\documentclass[preprint]{aastex}

\usepackage{emulateapj5}
\usepackage{apjfonts}

\newcommand{\ms}{\mbox{m\,s$^{-1}$}}
\newcommand{\bhyi}{\mbox{$\beta$~Hyi}}
\newcommand{\eboo}{\mbox{$\eta$~Boo}}

\newcommand{\Dnu}[1]{\Delta \nu_{#1}}
\newcommand{\half}{{\textstyle\frac{1}{2}}}
\newcommand{\muHz}{\mbox{$\mu$Hz}}
\let\epsilon\varepsilon

\slugcomment{Submitted to ApJ Letters}

\shorttitle{Oscillations in $\beta$ Hyi}
\shortauthors{Bedding et al.}

\begin{document}

\title{Evidence for solar-like oscillations in beta Hydri}

\author{
Timothy R. Bedding,\altaffilmark{1}
R. Paul Butler,\altaffilmark{2}
Hans Kjeldsen,\altaffilmark{3}
Ivan K. Baldry,\altaffilmark{4}\\
Simon J. O'Toole,\altaffilmark{1}
Christopher G. Tinney,\altaffilmark{4}
Geoffrey W. Marcy,\altaffilmark{5}
Francesco Kienzle\altaffilmark{6} and
Fabien Carrier\altaffilmark{6} 
}

\altaffiltext{1}{School of Physics A28, University of Sydney, NSW 2006,
Australia; bedding@physics.usyd.edu.au}

\altaffiltext{2}{Carnegie Institution of Washington,
Department of Terrestrial Magnetism, 5241 Broad Branch Road NW, Washington,
DC 20015-1305. }

\altaffiltext{3}{Theoretical Astrophysics Center, University of Aarhus,
DK-8000 Aarhus C, Denmark.}

\altaffiltext{4}{Anglo-Australian Observatory, P.O.\,Box 296, Epping, NSW
1710, Australia.}

\altaffiltext{5}{Department of Astronomy, University of California,
Berkeley, CA 94720; and Department of Physics and Astronomy, San Francisco,
CA 94132. }

\altaffiltext{6}{Observatoire de Gen\`eve, Ch.~des Maillettes 51, CH-1290
Sauverny, Switzerland.}

\begin{abstract} 
We have made a clear detection of excess power, providing strong evidence
for solar-like oscillations in the G2 subgiant \bhyi.  We observed this
star over five nights with the UCLES echelle spectrograph on the 3.9-m
Anglo-Australian Telescope, using an iodine absorption cell as a velocity
reference.  The time series of 1196 velocity measurements shows an rms
scatter of 3.30\,\ms, and the mean noise level in the amplitude spectrum at
frequencies above 0.5\,mHz is 0.11\,\ms.  We see a clear excess of power
centred at 1.0\,mHz, with peak amplitudes of about 0.5\,\ms, in agreement
with expectations for this star.  Fitting the asymptotic relation to the
power spectrum indicates the most likely value for the large separation is
56.2\,\muHz, also in good agreement with the known properties of \bhyi.
\end{abstract}

\keywords{stars: individual ($\beta$~Hyi) --- stars: oscillations---
techniques: radial velocities}

\section{Introduction}

The search for solar-like oscillations in other stars has been long and
difficult.  Observers have mostly concentrated on three stars: Procyon
($\alpha$~CMi), $\eta$~Boo and $\alpha$~Cen~A.  Reviews of those efforts
have been given by \citet{B+G94}, \citet[][hereafter KB95]{K+B95},
\citet{HJLaB96} and \citet{B+K98}.  More recently, \citet{KBF99} measured
Balmer-line equivalent widths in $\alpha$~Cen~A and set an upper limit on
oscillation amplitudes of only 1.4 times solar, with tentative evidence for
p-mode structure.  More recently still, measurements of velocity variations
in Procyon by \citet[][see also \citealp{BMM99}]{MSL99}, showed very good
evidence for oscillations with peak amplitudes of about 0.5\,\ms and
frequencies centred at about 1\,mHz.

Here, we report the clear detection of excess power, providing evidence for
oscillations in the G2 subgiant $\beta$~Hydri (HR~98, $V=2.80$, G2~IV).
This star is the closest G-type subgiant, with luminosity 3.5\,$L_\sun$,
mass 1.1\,$M_\sun$ and age about 6.7\,Gy \citep*{DLvdB98}.  It has received
less attention than the three stars mentioned above, presumably due to its
extreme southerly declination ($-77\degr$).  An attempt to measure
oscillations in \bhyi{} in radial velocity was made by \citet{E+C95} and
gave upper limits on the strongest modes of 1.5 to 2.0\,\ms, consistent
with our detection.

\section{Observations and data reduction}

The observations were made over five nights (2000 June 11--15).  We used
the University College London Echelle Spectrograph (UCLES) at the coud\'e
focus of the 3.9-m Anglo-Australian Telescope (AAT) at Siding Spring
Observatory, Australia.  To produce high-precision velocity measurements,
the star was observed through an iodine absorption cell mounted directly in
the telescope beam, immediately behind the spectrograph entrance slit.  The
cell is temperature-stabilized at 55$\pm$0.1$^\circ$C and imprints a rich
forest of molecular iodine absorption lines from 500\,nm to 600\,nm
directly on the incident starlight.  Echelle spectra were recorded with the
MITLL 2k$\times$4k 15$\mu$m pixel CCD, denoted MITLL2a, which covered the
wavelength range 470--880\,nm.

Exposure times were typically 60\,s, with a dead-time of 55\,s between
exposures (using ``FAST'' readout).  The signal-to-noise ratio for most
spectra was in the range 200 to 400, depending on the seeing and
extinction.  In total, 1196 spectra were collected, with the following
distribution over the five nights: 59, 166, 301, 325 and 345.  The first
two nights were affected by poor weather and technical problems.

Extraction of radial velocities from the echelle spectra followed the
method described by \citet{BMW96}.  This involved using the embedded iodine
lines both as a wavelength reference and also to recover the spectrograph
point-spread-function.  Essential to this process were template spectra
taken of \bhyi{} with the iodine cell removed from the beam, and of the
iodine cell itself superimposed on a rapidly rotating B-type star.

\begin{figure*}
\epsscale{0.75} 
\plotone{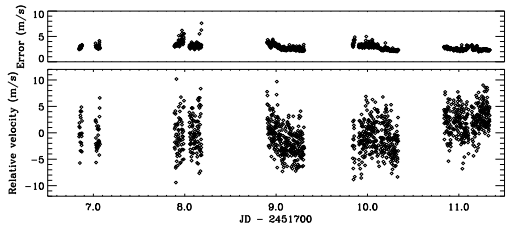}
\caption[]{\label{fig.ucles-time}Velocity measurements of \bhyi{}
obtained with the AAT (lower panel) and the corresponding uncertainties
(upper panel).  }
\end{figure*}

The resulting velocity measurements for \bhyi{} are shown in the lower
panel of Fig.~\ref{fig.ucles-time}.  They have been corrected to the solar
system barycentre, as described by \citet{BMW96}.  No other corrections,
decorrelation or high-pass filtering have been applied: the measurements
are exactly as they emerged from the pipeline processing.  Note that the
velocity measurements are relative to the velocity of the star when the
template was taken.  The slow variations in velocity within and between
nights are due to uncorrected instrumental drifts.  The rms scatter of
these measurements is 3.30\,\ms.

Uncertainties for the velocity measurements were estimated from residuals
in the fitting procedure and are shown in the upper panel of
Fig.~\ref{fig.ucles-time}.  Most lie in the range 2.5--4\,\ms, and fall
gradually during each night as the target rises in the sky.

Observations of \bhyi{} were also made using the CORALIE spectrograph on
the 1.2-m Leonard Euler Swiss telescope at La Silla Observatory in Chile
\citep{QMW2000}.  The precision of those measurements was poorer than
those from UCLES, presumably due -- at least in part -- to the smaller
telescope aperture.  The CORALIE data are not included in this Letter;
their analysis is postponed to a future paper.

\section{Time series analysis}	\label{sec.time-series}

\begin{figure*}
\epsscale{0.8} 
\plotone{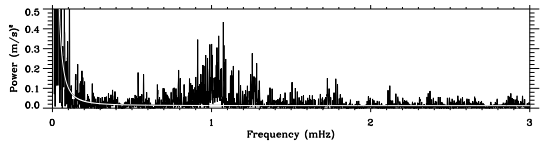}
\caption[]{\label{fig.ucles-power} Power spectrum of the AAT velocity
measurement of \bhyi{}.  The white line shows a two-component noise model
(see Sec.~\ref{sec.time-series}).}
\end{figure*}

The amplitude spectrum of the velocity time series was calculated as a
weighted least-squares fit of sinusoids \citep{FJK95,AKN98}, with a weight
being assigned to each point according its uncertainty estimate.
Figure~\ref{fig.ucles-power} shows the resulting power spectrum.  There is
a striking excess of power around 1\,mHz which is the clear signature of
solar-like oscillations.  As discussed below, the frequency and amplitude
of this excess power are in excellent agreement with expectations.  We also
note that the excess is apparent in the power spectra of individual nights.

Typically for such a power spectrum, the noise has two components: 
\begin{enumerate}

\item At high frequencies it is flat (i.e., white), indicative of the
Poisson statistics of photon noise.  The mean noise level in the amplitude
spectrum in the range 3--5\,mHz is 0.11\,\ms.  Since this is based on 1196
measurements, we can calculate (e.g., KB95)
that the velocity precision on the corresponding timescales is 2.2\,\ms.
%
%

\item Towards the lowest frequencies, we see rising power that arises from
the slow drifts mentioned above.  A logarithmic plot shows that, as
expected for instrumental drift, this noise goes inversely with frequency
in the amplitude spectrum (and inversely with frequency squared in power).
It has a value at 0.1\,mHz of 0.3\,\ms.
\end{enumerate}
The white line in Fig.~\ref{fig.ucles-power} shows the combined noise
level, and we see that the contribution from $1/f$ noise is neglible above
about 0.5\,mHz.

\section{Discussion}

\subsection{Oscillation frequencies}

Mode frequencies for low-degree oscillations in the Sun are reasonably well
approximated by the asymptotic relation:
\begin{equation}
  \nu(n,l) = \Dnu{} (n + \half l + \epsilon) - l(l+1) D_0.
        \label{eq.asymptotic}
\end{equation}
Here $n$ and $l$ are integers which define the radial order and angular
degree of the mode, respectively; $\Dnu{}$ (the so-called large separation)
reflects the average stellar density, $D_0$ is sensitive to the sound speed
near the core and $\epsilon$ is sensitive to the surface layers.  Note that
$D_0$ is $\delta\nu_0/6$, where $\delta\nu_0$ is the so-called small
frequency separation between adjacent modes with $l=0$ and $l=2$.

A similar relation to equation~(\ref{eq.asymptotic}) is expected for other
solar-like stars, although there may be significant deviations in more
evolved stars.  Models of the G~subgiant \eboo{} suggest that modes with
$l=1$ undergo `avoided crossings,' in which their frequencies are shifted
from their usual regular spacing by effects of gravity modes in the stellar
core \citep*{ChDBK95}.  Some evidence for this effect was seen in the
proposed detection of oscillations in \eboo{} by \citet{KBV95}.

In attempting to find peaks in our power spectrum matching the asymptotic
relation, we were severely hampered by the single-site window function.  As
is well known, daily gaps in a time series produce aliases in the power
spectrum at spacings $\pm11.57\,\muHz$ which are difficult to disentangle
from the genuine peaks.  Various methods have been discussed in the
literature for searching for a regular series of peaks, such as
autocorrelation, comb response and histograms of frequencies
\citep[e.g.,][]{GBK93,KBV95,MMM98,BMM99,MSL99}.  Here we use a type of comb
analysis, in which we calculated a response function for all sensible
values of $\Dnu{}$, $D_0$ and $\epsilon$.

Since we are searching for mode structure in the region of excess power, it
is convenient to rewrite equation~(\ref{eq.asymptotic}) as
\begin{equation}
  \nu(n',l) = \nu_0 + \Dnu{} (n' + \half l) - l(l+1) D_0.
        \label{eq.asymptotic-mod}
\end{equation}
Here, $\nu_0$ is the frequency of a radial ($l=0$) mode in the region of
maximum power, for which $n'=0$.  Thus, $\nu_0$ replaces $\epsilon$ as the
parameter for the absolute position of the comb.

The comb response function was obtained as follows.  We first thresholded
the power spectrum at the noise level.  In other words, any points less
than the noise level were set at this level.  For each triplet ($\Dnu{}$,
$D_0$, $\nu_0$) we then measured the (modified) power spectrum at each of
the frequencies predicted by equation~(\ref{eq.asymptotic-mod}).  We summed
these numbers to produce a response value which is a measure of the
goodness-of-fit of that triplet.  We limited the sum to $l=0,1,2$, and $n'
= -4, -3, \ldots, 4$, with $\nu_0$ centred at 1000\,\muHz.  The restriction
on $n'$ was made for two reasons: as in the Sun, we only see modes excited
to observable amplitudes over a restricted range of frequencies and, also
as in the Sun, we expect deviations from equation~(\ref{eq.asymptotic-mod})
over large ranges in frequency.

\begin{figure*}
\epsscale{0.85} 
\plotone{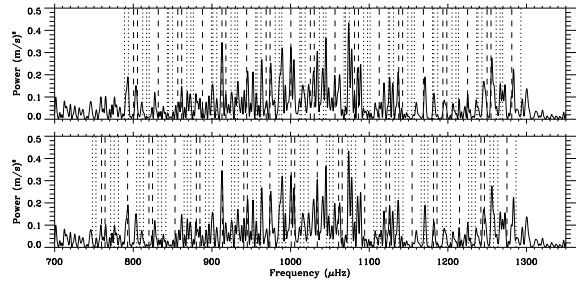}
\caption[]{\label{fig.ucles-freq}Close-up of the power spectrum of \bhyi,
with dashed lines showing the frequencies given by
equation~(\ref{eq.asymptotic-mod}) for solutions A (upper panel) and B~(lower
panel).  The dotted lines show the $\pm$1/day aliases of those
frequencies. }
\end{figure*}

{}From this analysis we identified the parameters that maximised the comb
response, searching over the ranges $\Dnu{}$ = 45--80\,\muHz{} and $D_0$ =
0.5--2.0\,\muHz.  These ranges encompass the values that would be expected
for this star.  For $\nu_0$, we examined a range of width $\Dnu{}$ centred
on 1000\,\muHz.  The following two triplets of ($\Dnu{}$, $D_0$, $\nu_0$)
gave the best responses: (56.2, 0.83, 1030.1)\,\muHz{} (solution~A) and
(60.3, 0.75, 1005.3)\,\muHz{} (solution~B).  The frequencies corresponding
to these solutions are shown in Fig.~\ref{fig.ucles-freq}, overlaid on the
measured power spectrum.  The distribution as a function of frequency of
matches between calculated and observed peaks is not the same for the two
solutions.  The matches in Solution~B occur mostly at the lower
frequencies, while those for Solution~A cover the full range of excess
power.  This suggests that Solution~A is more likely to be the correct one.

How do these two solutions agree with expectations for \bhyi?  The large
separation $\Dnu{}$ of a star scales approximately as the square root of
density.  Extrapolating from the solar case ($\Dnu{}=135.0\,\muHz$) using
parameters for \bhyi{} of $L=3.5\,L_\sun$, $T_{\rm eff}=5800$\,K and $M=
1.1\,M_\sun$ \citep[see][and references therein]{DLvdB98} gives
$\Dnu{}=56.0\,\muHz$.  Given that we searched over the entire range
45--80\,\muHz, the excellent agreement between observation and theory is
very encouraging, especially for solution~A.

There is no simple way to estimate $D_0$ for a star as evolved as \bhyi{},
since this parameter is very sensitive to evolution.  Extrapolating the
grid calculated by \citet{ChD93b} leads us to expect $D_0$ in the range 0.5
to 1.0, which is consistent with both solutions.  The value of $\nu_0$ (or,
equivalently,~$\epsilon$) depends on details of the surface layers, which
also requires construction of models specifically for \bhyi.

We note that simulations in which input frequencies obeyed the asymptotic
relation precisely showed a more definite (and unambiguous) peak in the
comb response than did the real data.  We conclude that the oscillation
frequencies in \bhyi{} have significant departures from the asymptotic
relation.  This is of great astrophysical interest and not unexpected, but
it does make it very difficult to extract the correct frequencies from our
single-site data.

At the suggestion of the referee, we examined the comb response of random
power spectra that were generated by multiplying white noise by an envelope
similar to the observed power excess.  For these spectra, the comb response
sometimes showed peaks similar to those from the real data, at various
values of~$\Delta\nu$.  The comb analysis therefore does not prove that the
excess power is due to individual modes.  However, given (i)~the agreement
between observations and expectations in all respects -- position and
amplitude of excess power (see below), and the best-fit value for
~$\Delta\nu$ -- and (ii)~the absence of any theoretical or observational
reason to attribute the power excess to other sources, we consider that
solar-like oscillations are the most likely explanation for the observed
power excess.

\subsection{Oscillation amplitudes}

The strongest peaks in the amplitude spectrum of \bhyi{} (square root of
power) reach about 0.6\,\ms.  However, these are likely to have been
strengthened significantly by constructive interference with noise peaks.
As stressed by 
KB95 (Appendix A.2), the effects of the noise must
be taken into account when estimating the amplitude of the underlying
signal.  To do this, we have generated simulated time series consisting of
artificial signal plus noise.  We conclude that the underlying oscillations
have peaks of about 0.5\,\ms.

Solar-like oscillations are excited by convection and the expected
amplitudes have been estimated using theoretical models.  Based on models
by \citet{ChD+F83}, KB95
suggested that amplitudes in velocity should scale as $L/M$.  More recent
calculations by \citet{HBChD99} confirm this scaling relation, at least for
stars with near-solar effective temperatures.  For \bhyi{}, the implied
amplitude is about 3.2 times solar, which is 0.7 to 0.8\,\ms{} (based on
the strongest few peaks -- see KB95).  The observed amplitudes are
therefore consistent with expectations.  We also note that the frequency of
excess power (1\,mHz) is in excellent agreement with the value expected
from scaling the acoustic cutoff from the solar case \citep[][KB95]{BGN91}.

\section{Conclusion}

Our observations of \bhyi{} show an obvious excess of power, clearly
separated from the $1/f$ noise, and with a position and amplitude that are
in agreement with expectations.  Although hampered by the single-site
window, a comb analysis shows evidence for approximate regularity in the
peaks at the spacing expected from asymptotic theory.  There seem to be
significant departures from regularity, perhaps indicating mode shifts from
avoided crossings.  We hope that further analysis along one or more of the
following lines will allow us to explore further the oscillation spectrum
of \bhyi: including the CORALIE observations which, despite their lower
precision, could usefully improve the spectral window; reducing the noise
of the velocity measurements (UCLES and CORALIE) by decorrelating against
external parameters \citep[e.g.,][]{GBD91}; combining these results with an
analysis of equivalent-width variations of strong lines in the spectral
region uncontaminated by iodine lines (e.g., H$\alpha$); and using
theoretical models to calculate expected shifts due to avoided crossings,
to help identify the affected modes.

The clear strength of the technique used in these observations is its
ability to precisely calibrate spectrograph variations at the timescales of
primary interest.  Although the best radial velocity precisions achieved in
the {\em long term\/} from this technique are 3--4\,\ms, our data clearly
demonstrate that the precision at frequencies around 1\,mHz is
significantly better (2.2\,\ms).  Our results provide valuable confirmation
that oscillations in solar-like stars really do have the amplitudes that we
have been led to expect by extrapolating from the Sun.  This bodes
extremely well for success of space missions such as MOST \citep{MKW2000},
MONS \citep*{KBChD2000} and COROT \citep{Bag98}, which will provide
photometric data of high quality for a wide sample of stars.

\acknowledgments

This work was supported financially by: the Australian Research Council
(TRB and SJOT), NSF grant AST-9988087 (RPB), SUN Microsystems, and the
Danish Natural Science Research Council and the Danish National Research
Foundation through its establishment of the Theoretical Astrophysics Center
(HK).


\begin{thebibliography}{25}
\expandafter\ifx\csname natexlab\endcsname\relax\def\natexlab#1{#1}\fi
\expandafter\ifx\csname url\endcsname\relax
  \def\url#1{{\tt #1}}\fi

\bibitem[Arentoft et~al.(1998)Arentoft, Kjeldsen, Nuspl, Bedding, Fronto,
  Viskum, Frandsen, \& Belmonte]{AKN98}
Arentoft, T., Kjeldsen, H., Nuspl, J., Bedding, T.~R., Fronto, A., Viskum, M.,
  Frandsen, S., \& Belmonte, J.~A., 1998, A\&A, 338, 909.

\bibitem[Baglin et~al.(1998)]{Bag98}
Baglin, A., et~al., 1998, In Deubner, F.-L., Christensen-Dalsgaard, J., \&
  Kurtz, D.~W., editors, {\em Proc.\ IAU Symp.\ 185, New Eyes to See Inside the
  Sun and Stars}, page 301. Dordrecht: Kluwer.
\newblock see also {\tt
  http://\linebreak[0]www.astrsp-mrs.fr/\linebreak[0]projets/corot/}.

\bibitem[Barban et~al.(1999)Barban, {Michel}, {Martic}, {Schmitt}, {Lebrun},
  {Baglin}, \& {Bertaux}]{BMM99}
Barban, C., {Michel}, E., {Martic}, M., {Schmitt}, J., {Lebrun}, J.~C.,
  {Baglin}, A., \& {Bertaux}, J.~L., 1999, A\&A, 350, 617.

\bibitem[Bedding \& Kjeldsen(1998)]{B+K98}
Bedding, T.~R., \& Kjeldsen, H., 1998, In Donahue, R.~A., \& Bookbinder, J.~A.,
  editors, {\em Tenth Cambridge Workshop on Cool Stars, Stellar Systems and the
  Sun}, volume 154 of {\em A.S.P. Conf.\ Ser.}, page 301. San Francisco: ASP.

\bibitem[Brown \& Gilliland(1994)]{B+G94}
Brown, T.~M., \& Gilliland, R.~L., 1994, ARA\&A, 33, 37.

\bibitem[Brown et~al.(1991)Brown, Gilliland, Noyes, \& Ramsey]{BGN91}
Brown, T.~M., Gilliland, R.~L., Noyes, R.~W., \& Ramsey, L.~W., 1991, ApJ, 368,
  599.

\bibitem[Butler et~al.(1996)Butler, Marcy, Williams, McCarthy, Dosanjh, \&
  Vogt]{BMW96}
Butler, R.~P., Marcy, G.~W., Williams, E., McCarthy, C., Dosanjh, P., \& Vogt,
  S.~S., 1996, PASP, 108, 500.

\bibitem[Christensen-Dalsgaard(1993)]{ChD93b}
Christensen-Dalsgaard, J., 1993, In Brown, T.~M., editor, {\em GONG~1992:
  Seismic Investigation of the Sun and Stars}, volume~42 of {\em A.S.P. Conf.\
  Ser.}, page 347. Utah: Brigham Young.

\bibitem[Christensen-Dalsgaard et~al.(1995)Christensen-Dalsgaard, Bedding, \&
  Kjeldsen]{ChDBK95}
Christensen-Dalsgaard, J., Bedding, T.~R., \& Kjeldsen, H., 1995, ApJ, 443,
  L29.

\bibitem[Christensen-Dalsgaard \& Frandsen(1983)]{ChD+F83}
Christensen-Dalsgaard, J., \& Frandsen, S., 1983, Sol. Phys., 82, 469.

\bibitem[Dravins et~al.(1998)Dravins, Lindegren, \& VandenBerg]{DLvdB98}
Dravins, D., Lindegren, L., \& VandenBerg, D.~A., 1998, A\&A, 330, 1077.

\bibitem[Edmonds \& Cram(1995)]{E+C95}
Edmonds, P.~D., \& Cram, L.~E., 1995, MNRAS, 276, 1295.

\bibitem[Frandsen et~al.(1995)Frandsen, Jones, Kjeldsen, Viskum, Hjorth,
  Andersen, \& Thomsen]{FJK95}
Frandsen, S., Jones, A., Kjeldsen, H., Viskum, M., Hjorth, J., Andersen, N.~H.,
  \& Thomsen, B., 1995, A\&A, 301, 123.

\bibitem[Gilliland et~al.(1991)Gilliland, Brown, Ducan, Suntzeff, Lockwood,
  Thompson, Schild, Jeffrey, \& Penprase]{GBD91}
Gilliland, R.~L., Brown, T.~M., Ducan, D.~K., Suntzeff, N.~B., Lockwood, G.~W.,
  Thompson, D.~T., Schild, R.~E., Jeffrey, W.~A., \& Penprase, B.~E., 1991, AJ,
  101, 541.

\bibitem[Gilliland et~al.(1993)Gilliland, Brown, Kjeldsen, McCarthy, Peri,
  et~al.]{GBK93}
Gilliland, R.~L., Brown, T.~M., Kjeldsen, H., McCarthy, J.~K., Peri, M.~L.,
  et~al., 1993, AJ, 106, 2441.

\bibitem[Heasley et~al.(1996)Heasley, Janes, Labonte, Guenther, Mickey, \&
  Demarque]{HJLaB96}
Heasley, J.~N., Janes, K., Labonte, B., Guenther, D., Mickey, D., \& Demarque,
  P., 1996, PASP, 108, 385.

\bibitem[Houdek et~al.(1999)Houdek, {Balmforth}, {Christensen-Dalsgaard}, \&
  {Gough}]{HBChD99}
Houdek, G., {Balmforth}, N.~J., {Christensen-Dalsgaard}, J., \& {Gough}, D.~O.,
  1999, A\&A, 351, 582.

\bibitem[Kjeldsen \& Bedding(1995)]{K+B95}
Kjeldsen, H., \& Bedding, T.~R., 1995, A\&A, 293, 87 (KB95).

\bibitem[Kjeldsen et~al.(2000)Kjeldsen, Bedding, \&
  Christensen-Dalsgaard]{KBChD2000}
Kjeldsen, H., Bedding, T.~R., \& Christensen-Dalsgaard, J., 2000, In Szabados,
  L., \& Kurtz, D., editors, {\em IAU Colloqium 176: The Impact of Large-Scale
  Surveys on Pulsating Star Research}, volume 203, page~73. ASP Conf. Ser.
\newblock see also {\tt http://\linebreak[0]astro.ifa.au.dk/\linebreak[0]MONS}.

\bibitem[Kjeldsen et~al.(1999)Kjeldsen, Bedding, Frandsen, \& Dall]{KBF99}
Kjeldsen, H., Bedding, T.~R., Frandsen, S., \& Dall, T.~H., 1999, MNRAS, 303,
  579.

\bibitem[Kjeldsen et~al.(1995)Kjeldsen, Bedding, Viskum, \& Frandsen]{KBV95}
Kjeldsen, H., Bedding, T.~R., Viskum, M., \& Frandsen, S., 1995, AJ, 109, 1313.

\bibitem[Martic et~al.(1999)Martic, Schmitt, Lebrun, Barban, Connes, Bouchy,
  Michel, Baglin, Appourchaux, \& Bertaux]{MSL99}
Martic, M., Schmitt, J., Lebrun, J.-C., Barban, C., Connes, P., Bouchy, F.,
  Michel, E., Baglin, A., Appourchaux, T., \& Bertaux, J.-L., 1999, A\&A, 351,
  993.

\bibitem[Matthews et~al.(2000)Matthews, Kuschnig, Walker, {Pazder}, {Johnson},
  {Skaret}, {Shkolnik}, {Lanting}, {Morgan}, \& {Sidhu}]{MKW2000}
Matthews, J.~M., Kuschnig, R., Walker, G. A.~H., {Pazder}, J., {Johnson}, R.,
  {Skaret}, K., {Shkolnik}, E., {Lanting}, T., {Morgan}, J.~P., \& {Sidhu}, S.,
  2000, In Szabados, L., \& Kurtz, D., editors, {\em IAU Colloqium 176: The
  Impact of Large-Scale Surveys on Pulsating Star Research}, volume 203,
  page~74. ASP Conf. Ser.
\newblock see also {\tt
  http://\linebreak[0]www.astro.ubc.ca/\linebreak[0]MOST}.

\bibitem[Mosser et~al.(1998)Mosser, {Maillard}, {M\'ekarnia}, \& {Gay}]{MMM98}
Mosser, B., {Maillard}, J.~P., {M\'ekarnia}, D., \& {Gay}, J., 1998, A\&A, 340,
  457.

\bibitem[Queloz et~al.(2000)Queloz, {Mayor}, {Weber}, {Bl{\'e}cha}, {Burnet},
  {Confino}, {Naef}, {Pepe}, {Santos}, \& {Udry}]{QMW2000}
Queloz, D., {Mayor}, M., {Weber}, L., {Bl{\'e}cha}, A., {Burnet}, M.,
  {Confino}, B., {Naef}, D., {Pepe}, F., {Santos}, N., \& {Udry}, S., 2000,
  A\&A, 354, 99.

\end{thebibliography}
\end{document}